\documentclass[lettersize,journal]{IEEEtran}
\usepackage{amsmath,amsfonts}
\usepackage{array}
\usepackage{textcomp}
\usepackage{stfloats}
\usepackage{url}
\usepackage{verbatim}
\usepackage{graphicx}
\usepackage{cite}
\usepackage{booktabs} 
\usepackage{mleftright}
\mleftright
\usepackage{algorithm}  
\usepackage{algorithmicx}  
\usepackage{algpseudocode}
\usepackage[caption=false,font=footnotesize]{subfig}

\usepackage{diagbox}
\usepackage{graphicx}
\usepackage{multirow}
\usepackage{makecell}

\def\D {\mathcal{D}}

\def\S {\mathcal{S}}

\renewcommand{\thetable}{\arabic{table}}

\usepackage{amsmath,amsfonts,bm}
\def\cleanone{\textit{clean-up-\uppercase\expandafter{\romannumeral1}}}
\def\cleantwo{\textit{clean-up-\uppercase\expandafter{\romannumeral2}}}

\def\eqref#1{equation~\ref{#1}}

\def\1{\bm{1}}

\def\va{{\bm{a}}}

\def\ve{{\bm{e}}}
\def\vf{{\bm{f}}}

\def\vm{{\bm{m}}}

\def\vs{{\bm{s}}}

\def\vv{{\bm{v}}}

\def\vx{{\bm{x}}}

\DeclareMathAlphabet{\mathsfit}{\encodingdefault}{\sfdefault}{m}{sl}
\SetMathAlphabet{\mathsfit}{bold}{\encodingdefault}{\sfdefault}{bx}{n}

\def\gE{{\mathcal{E}}}

\def\gG{{\mathcal{G}}}

\def\gV{{\mathcal{V}}}

\newtheorem{defn}{Definition}

\begin{document}

\title{GRLinQ: An Intelligent Spectrum Sharing Mechanism for Device-to-Device Communications with Graph Reinforcement Learning}

\author{
\IEEEauthorblockN{Zhiwei Shan,
                Xinping Yi,
                Le Liang, 
                    Chung-Shou Liao,
                    and Shi Jin}

                    
                    
                     

\thanks{This work was presented in part at IEEE International Symposium on Information Theory, Athens, Greece, 2024.}
\thanks{Z.~Shan is with University of Liverpool, Liverpool L69 3GJ, United Kingdom, and also with National Tsing Hua University, Hsinchu 30013, Taiwan, ROC. Email: zshan@liverpool.ac.uk.}
\thanks{X.~Yi, L.~Liang, and S.~Jin are with National Mobile Communications Research Laboratory, Southeast University, Nanjing 210096, China. L. Liang is also with Purple Mountain Laboratories, Nanjing 211111, China. Email: \{xyi, lliang, jinshi\}@seu.edu.cn.}
\thanks{C.-S.~Liao is with National Tsing Hua University, Hsinchu 30013, Taiwan, ROC. Email:                      csliao@ie.nthu.edu.tw.}
}

\maketitle

\begin{abstract}
Device-to-device (D2D) spectrum sharing in wireless communications is a challenging non-convex combinatorial optimization problem, involving entangled link scheduling and power control in a large-scale network. The state-of-the-art methods, either from a model-based or a data-driven perspective, exhibit certain limitations such as the critical need for channel state information (CSI) and/or a large number of (solved) instances (e.g., network layouts) as training samples. To advance this line of research, we propose a novel hybrid model/data-driven spectrum sharing mechanism with graph reinforcement learning for link scheduling (GRLinQ), injecting information theoretical insights into machine learning models, in such a way that link scheduling and power control can be solved in an intelligent yet explainable manner. 
Through an extensive set of experiments, GRLinQ demonstrates superior performance to the existing model-based and data-driven link scheduling and/or power control methods, with a relaxed requirement for CSI, a substantially reduced number of unsolved instances as training samples, a possible distributed deployment, reduced online/offline computational complexity, and more remarkably excellent scalability and generalizability over different network scenarios and system configurations.

\end{abstract}

\section{Introduction}
\label{sec:Introduction}
Device-to-device (D2D) communications are going to play a central role in future wireless networks, enabling direct message exchange between devices without resorting to cellular infrastructure. 
From an information theoretical perspective, the underlying physical model of D2D communications is a Gaussian interference channel (GIC), in which the capacity characterization has been a long-standing open problem.
Among various sophisticated coding techniques for GIC, treating interference as noise (TIN) has been regarded as a low-complexity and robust one deployable in large-scale D2D networks. Remarkably, it has been proven that, with proper link scheduling, for a selected subset of D2D links with certain conditions satisfied, TIN at the receivers and power control at the transmitters can achieve the information-theoretical optimality in an asymptotic sense \cite{geng2015optimality,geng2016optimality,geng2015optimality2,sun2016optimality,gherekhloo2015sub,joudeh2019optimality,yi2019opportunistic,joudeh2020optimality,gherekhloo2016expanded}.

Nevertheless, D2D link scheduling and power control is still a challenging problem, aiming to maximize overall network throughput by activating a subset of D2D links with proper power allocation for each link. The primary challenge arises from the entangled nature of link scheduling and power control, where the former is a non-convex combinatorial optimization problem, particularly formidable in large-scale networks, and the latter can be seen as a continuous version of the former, yet optimizing the latter is critically dependent on the former. 
Due to the complex entanglement of link scheduling and power control, conventional wisdom suggests tackling the two problems separately, yielding a number of effective and low-complexity methods in the literature.
%

Current state-of-the-art link scheduling methods include traditional model-based approaches and emerging data-driven techniques. Model-based methods such as FlashLinQ\cite{Flash}, ITLinQ\cite{IT}, and ITLinQ+\cite{IT+} design sequential link selection algorithms from an information-theoretical perspective. However, the adjustable parameters they use rely highly on network scenarios and system parameters. It is time/effort-consuming to find appropriate values for each different network scenario, while the same set of parameters apparently does not work across all networks. FPLinQ \cite{FPLinQ} approaches the optimization problem from a fractional programming standpoint. It relaxes the requirements of tuning parameters but requires centralized processing with accurate CSI, which is difficult to obtain in large-scale networks, especially in a distributed setting.
When relaxing the CSI requirement, data-driven approaches hinge on the power of deep neural networks (DNNs), leaving aside insights obtained from model-based methods. As such, SpatialLinQ \cite{spatial} involves learning interference patterns among neighboring transmitters/receivers through kernels, and then generating outcomes through DNNs. However, it requires hundreds of thousands of solved instances as training samples, i.e., different network layouts and the corresponding link scheduling solutions by e.g., FPLinQ. A graph-embedding-based method, GELinQ \cite{GNNLinQ}, reduces this number to hundreds of training network layouts, though at the cost of degraded performance. Although promising, the existing data-driven approaches such as SpatialLinQ and GELinQ simply put network layout or geographical location as node features, without considering the insights obtained from the model-based methods, rendering degraded sum rate performance and/or reduced generalizability.

Current state-of-the-art power control methods encompass traditional model-based approaches, and model/data-driven deep learning approaches. The previously mentioned model-based ITLinQ+\cite{IT+} provides a power control solution together with link scheduling, yet its effectiveness relies highly on the exponentially decayed channel strength. While the optimization-based FPLinQ \cite{FPLinQ} offers an effective power control solution to practical channels, its centralized processing requirement and high computational complexity make it challenging to be deployed in practice.
The weighted minimum mean squared error (WMMSE) \cite{WMMSE} method focuses on minimizing a reformulated weighted mean-squared-error cost function, yielding superior sum rate performance. Although these methods can yield results very close to the optimum, they are often criticized for their high complexity and the need for accurate CSI acquisition \cite{shen2020graph,UWMMSE,hu2020iterative}. 
The model-driven method, unfolded WMMSE (UWMMSE) \cite{UWMMSE},  unrolls the WMMSE iterations into a cascade of DNN layers, each of which retains the same update structure as the original algorithm, but with parameters learned by graph neural networks (GNNs) from data. Compared to the number of iterations required by WMMSE, UWMMSE can use fewer layers to achieve similar results, thereby reducing computational complexity. PCGNN \cite{PCGNN}, as a data-driven method, utilizes a GNN to directly learn the power allocation function, achieving near-optimal performance with much fewer training samples. However, all the aforementioned power control methods rely on exact CSI and still have relatively high computational cost, i.e., quadratic complexity with respect to the number of D2D links.

\begin{table*}[tbp]
\caption{Summary of state-of-the-art link scheduling methods}
\label{summary_LS}
\resizebox{2\columnwidth}{!}{
\begin{tabular}{|c|c|c|c|c|c|}
\hline
Method                        & \makecell{FlashLinQ\cite{Flash},\\ITLinQ\cite{IT}, ITLinQ+\cite{IT+}}         & FPLinQ\cite{FPLinQ}                           & SpatialLinQ\cite{spatial}           & GELinQ\cite{GNNLinQ}                 & GRLinQ (this work)            \\ \hline\hline
Driving Mechanism             & Model-based                       & Model-based                     & Data-driven           & Data-driven            & Hybrid model/data-driven \\ \hline
Methodology                   & \makecell{Heuristic\\sequential selection}     & \makecell{Mathematical\\optimization} & Machine learning                    & Machine learning                     & Machine learning                \\ \hline
Key Idea                      & \makecell{TIN conditions} & \makecell{Fractional\\programming}           & Kernel \& DNN  & GNN \& DNN & RL \& GNN         \\ \hline
CSI                           & Yes                                & Yes                              & No                    & No                     & No                \\ \hline
System Architecture           & Decentralized                        & Centralized                      & Centralized           & Distributed            & Distributed       \\ \hline
\makecell{Number of Training\\Network Layouts} & /                                  & /                                & Hundreds of thousands & Hundreds               & Hundreds          \\ \hline
Generalizability              & /                                  & /                                & Strong                & Good                   & Strong            \\ \hline
Complexity                    & $O(N^2)$                           & $O(N^2)$                         & $O(N)$                & $O(N)$                 & $O(N)$               \\ \hline
\end{tabular}
}
\end{table*}

\begin{table*}[tbp]
\caption{Summary of state-of-the-art power control methods}
\label{summary_PC}
\resizebox{2\columnwidth}{!}{
\begin{tabular}{|c|c|c|c|c|c|}
\hline
Method & \makecell{WMMSE\cite{WMMSE}} & FPLinQ-pc\cite{FPLinQ} & UWMMSE\cite{UWMMSE} & PCGNN\cite{PCGNN}  & GRLinQ-pc (this work)           \\ \hline\hline
Driving Mechanism             & Model-based                       & Model-based                     & Model-driven           & Data-driven            & Hybrid model/data-driven \\ \hline
Methodology                   & \makecell{Mathematical\\optimization}     & \makecell{Mathematical\\optimization} & Machine learning                    & Machine learning                     & Machine learning                \\ \hline
Key Idea                      & \makecell{Minimum mean\\square error} & \makecell{Fractional\\programming}           & \makecell{Unfolding WMMSE \\ \& GNN}  & GNN & RL \& GNN         \\ \hline
CSI                           & Yes                                & Yes                              & Yes                    & Yes                     & No                \\ \hline
System Architecture           & Distributed                        & Centralized                      & Distributed           & Distributed            & Distributed       \\ \hline
\makecell{Number of Training\\Network Layouts} & /                                  & /                                & Hundreds of thousands & Tens of thousands & Hundreds  \\ \hline
Generalizability              & /                                  & /                                & Good                & Strong                   & Strong            \\ \hline
Complexity                    & $O(N^2)$                           & $O(N^2)$                         & $O(N^2)$               & $O(N^2)$                & $O(N)$               \\ \hline
\end{tabular}
}
\end{table*}

To advance this line of research, we propose a new hybrid model/data-driven intelligent spectrum sharing framework, named GRLinQ, addressing link scheduling and power control by leveraging advancements in graph reinforcement learning and some information theoretical insights.
Specifically,
we reformulate the link scheduling and power control problem as Markov decision processes (MDPs), breaking down the complex combinatorial optimization problem into multiple steps. By employing reinforcement learning (RL), we can learn and optimize the strategy for state transitions based on current network conditions. In doing so, it relaxes the need for labeled data, which is usually obtained from solved instances by e.g., FPLinQ. As a result, GRLinQ achieves more efficient resource utilization and improved overall performance, making it highly adaptable and effective in real-world scenarios.
Notably, GRLinQ employs GNNs as the policy and value networks within the RL framework, so as to effectively capture graph structures in interference graphs, demonstrating the significance and efficiency of topology awareness in both link scheduling and power control.
Another key enabling factor is that GRLinQ incorporates insights derived from model-based approaches, e.g., FlashLinQ, ITLinQ, and ITLinQ+. These insights are carefully integrated into the node features of the GNNs, enriching GRLinQ with a deeper understanding of network behaviors and performance measures.


Through an extensive set of experiments, GRLinQ demonstrates superior sum rate performance compared to the state-of-the-art link scheduling and power control benchmarks, excellent scalability across different network scenarios (e.g., network sizes, network density), and strong generalization performance (e.g., unseen carrier frequency, D2D link distance). 
In addition, GRLinQ also relaxes the requirements of accurate CSI and a large number of solved instances as training samples, and enables potential distributed deployment. 
The comparison with state-of-the-art methods is summarized in Table \ref{summary_LS} and \ref{summary_PC}. The proposed GRLinQ appears to be the most promising for dealing with D2D spectrum sharing with a hybrid model/data-driven machine learning methodology.
%
Specifically, GRLinQ possesses the following advantages over the state-of-the-art link scheduling and power control methods, according to the comprehensive experiments. 
\begin{itemize}
    \item For the link scheduling problem, GRLinQ achieves comparable sum rate performance as FPLinQ,
    significantly improved over FlashLinQ, ITLinQ, and ITLinQ+, while requiring only pairwise distance information rather than accurate CSI. Meanwhile, GRLinQ surpasses SpatialLinQ and GELinQ by a large margin, whilst eliminating the need of solved instances as training samples and reducing the amount of required training data.
    \item For the power control problem, GRLinQ-pc also approaches the best-known FPLinQ-pc and WMMSE for different network sizes, and strictly surpasses the existing model/data-driven deep learning approaches such as UWMMSE and PCGNN, while
    requiring only distance information rather than accurate CSI and substantially reduced training data. 
    \item
    Thanks to topology-awareness (i.e., $K$-nearest interference graphs) and information-theory-inspired feature design, GRLinQ possesses excellent scalability from trained smaller-sized to testing large-scale network scenarios and strong generalizability from training to testing under different unseen system configurations.
\end{itemize}

The rest of this paper is organized as follows. Section II describes wireless link scheduling and power control in D2D networks, and constructs $K$-nearest interference graphs. The existing link scheduling and power control methods are reviewed in Section III. The proposed GRLinQ framework is detailed in Section IV, and the extensive experiments are presented in Section V. We conclude the paper in Section VI.

\section{Problem Statement}\label{Problem Statement}
\subsection{Wireless Link Scheduling and Power Control}
Consider a D2D wireless network $\D$ with $N$ unicast links, denoted by $\D=\{D_i\}_{i=1}^N$. 
Each D2D link $D_i$ consists of a transmitter $\text{Tx}_i$ and a paired receiver $\text{Rx}_i$, each equipped with one single antenna. 
Modeled as a Gaussian interference channel, the received signal at Rx$_i$ can be written by
\begin{align}
    y_i = h_{ii}z_i + \sum_{j=1, j\neq i}^N h_{ji} z_j + n_i
\end{align}
where $h_{ji}$ is the channel coefficient between $\text{Tx}_j$ and $\text{Rx}_i$ integrating both large-scale fading (i.e., path-loss) and small-scale fading coefficients, $z_i$ is the transmitted signal from Tx$_i$ with the transmit power constraint $E[|z_i|^2] \leq p_i$, for all $i$,
and $n_i$ is the additive white Gaussian noise, i.e., $n_i \sim \mathcal{CN}(0,\sigma ^2)$. We consider link scheduling within a coherence block, where channel coefficients remain unchanged during communication.

\textbf{Link scheduling.}
Let us introduce $x_i \in \{0,1\}$ as an indicator variable for each link $D_i$, which equals $1$ if the link is scheduled and $0$ otherwise. 
Then, the achievable data rate of link $D_i$ when treating interference as noise is given by
\begin{IEEEeqnarray}{rCl}\label{Eq_rate}
R_i(\vx) = \log \left( 1+\frac{|h_{ii}|^2p_ix_i}{\sum_{j \neq i}|h_{ji}|^2p_jx_j+\sigma^2}\right),
\end{IEEEeqnarray}
where $\vx$ is a collection of $\{x_i\}_{i=1}^N$, and for convenience we assume full power transmission at the Txs. Due to mutual interference between links, activating too many links can lead to a decrease in data rate. Therefore, the task of wireless link scheduling is to activate a suitable set of links, aiming to maximize some utility function of the average rates.

In this paper, we formulate the link scheduling problem as maximizing the weighted sum rate over the $N$ links for each scheduling slot, i.e.,
\begin{IEEEeqnarray}{lll}
\label{obj_func}
   &\max_\vx \ \sum^N_{i=1}w_i R_i(\vx) \\
   &\text{subject to } x_i \in  \{0,1\},\ \forall i,\nonumber
\end{IEEEeqnarray}
where the weight $w_i$ signifies the priority assigned to link $D_i$. 

\textbf{Power control.}
To further enhance network performance, power control can be introduced as an additional mechanism. In this context, the indicator variable $x_i$ is relaxed to $x_i \in [0,1]$, allowing for a continuous range of power allocation levels for each transmitter $\text{Tx}_i$. This adjustment enables more fine-grained control over the transmit power, which can significantly improve the achievable data rates.


\subsection{Topology and $K$-nearest Interference Graphs}
\begin{figure}[htbp]
   \centering
   \vspace{-10pt}
   \subfloat[Topology graph.]{%
     \label{topo_graph}
     \includegraphics[width=3.5cm]{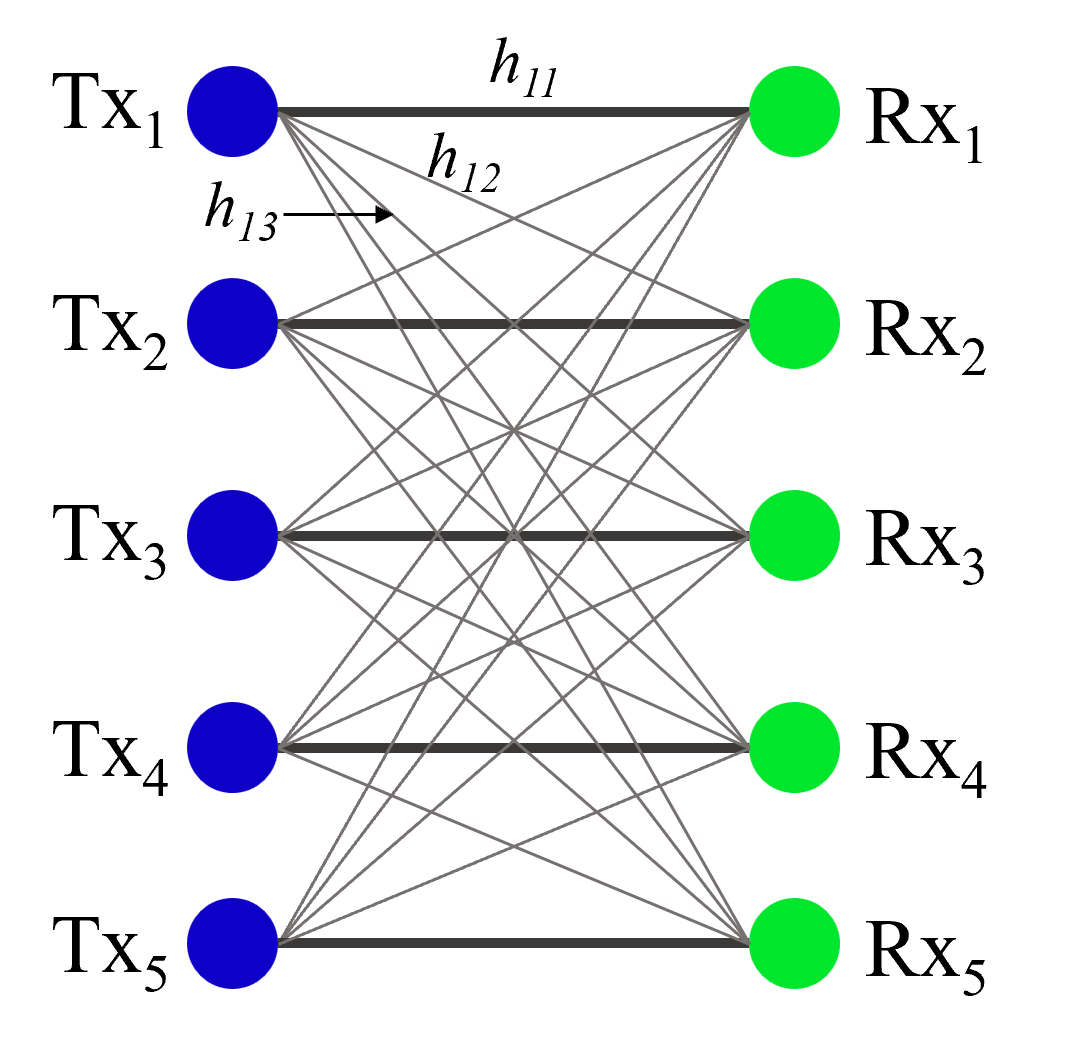}}
   \hfil
   \subfloat[$K$-nearest Interference graph.]{%
     \label{conf_graph}
     \includegraphics[width=4cm]{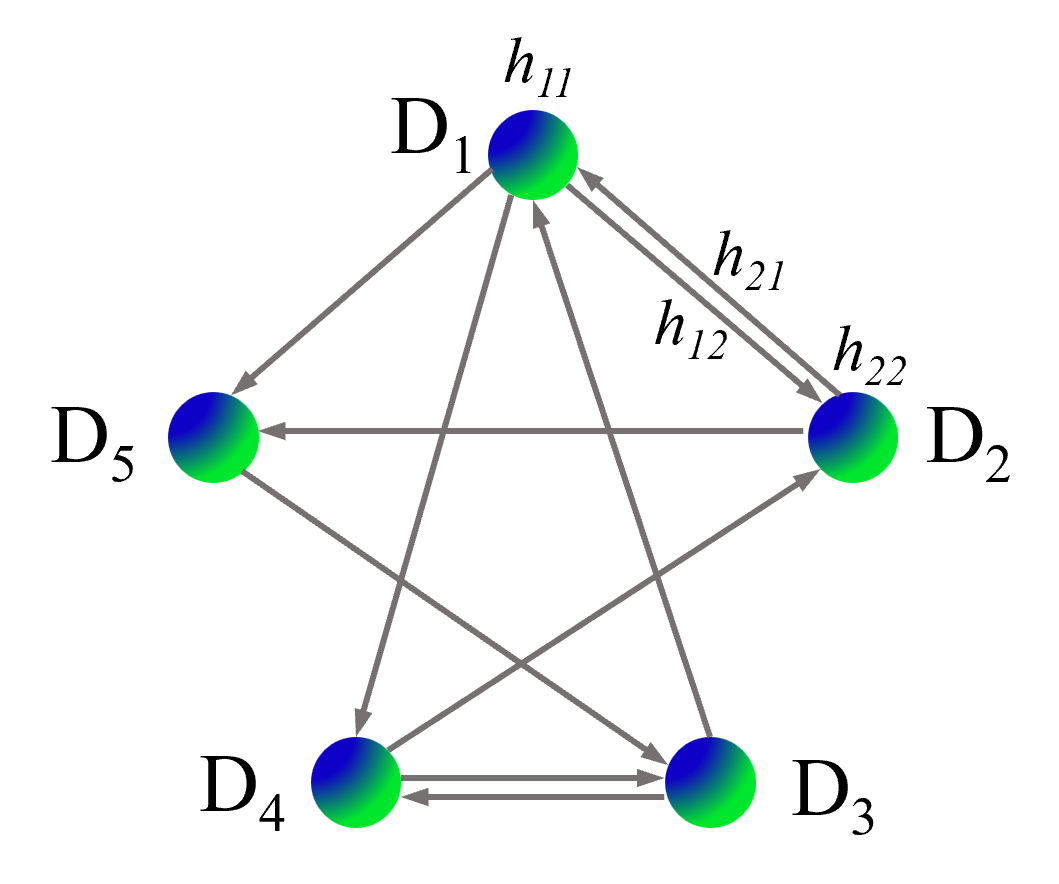}}  
   \caption{A D2D network with 5 links. (a) topology graph, and (b) the corresponding $K$-nearest Interference graph with $K=2$.}
   \label{topo_conf_graph}
\end{figure}

The D2D network can be represented by a topology graph, a bipartite graph that captures the Gaussian interference channel with edge weights being channel coefficients, as shown in Fig.\ref{topo_graph}. Each transmitter intends to send a message to its corresponding receiver, simultaneously causing interference to other receivers. 

The D2D network can also be represented by a $K$-nearest interference graph, 
as illustrated in Fig.\ref{conf_graph}. Such a graph captures the most significant interfering links of each receiver.

\begin{defn}[$K$-nearest Interference Graph] 
Given an $N$-link D2D network $\D$, the corresponding $K$-nearest interference graph is defined as a directed graph,
where each D2D link $D_i$ serves as a vertex, 
and the directed edge $(D_i, D_j)$ indicates
the interfering link from Tx$_i$ to Rx$_j$.
The number of incoming edges to any $D_i$ is $K$, i.e., $|\mathcal{N}(i)|=K \le N-1$, for all $i$, where $\mathcal{N}(i)$ represents the set of $K$ vertices exerting the most substantial influence to $D_i$, i.e., $|h_{ji}|^2 \ge |h_{ki}|^2$ for all $j \in \mathcal{N}(i)$ and $k \notin \mathcal{N}(i)$.
\end{defn}

The $K$-nearest interference graph imposes a structural property that can be exploited in algorithm designs. When $K = N - 1$, the $(N - 1)$-nearest interference graph becomes a fully connected directed graph, accounting for all interfering links. 


\section{The State-of-the-Art}\label{The State-of-the-Art}
\subsection{Link Scheduling}
The state-of-the-art link scheduling mechanisms include model-based approaches, such as FlashLinQ \cite{Flash}, ITLinQ \cite{IT}, ITLinQ+ \cite{IT+}, FPLinQ \cite{FPLinQ}, and data-driven approaches, such as SpatialLinQ \cite{spatial}, GELinQ \cite{GNNLinQ}, to name a few.  

\subsubsection{Model-based approaches}\label{Model-driven Approaches}
Model-based approaches rely on theoretical frameworks and predefined models for algorithm designs.
FlashLinQ, ITLinQ, and ITLinQ+ share a unified sequential selection framework. Denote the set of active links as $\S$, initially an empty set. The link scheduling mechanisms sequentially traverse all D2D links and evaluate each D2D link against predefined conditions. If a particular link satisfies the ``condition'', it is marked as active and included in $\S$; otherwise, it is labeled as inactive. FlashLinQ, ITLinQ, and ITLinQ+ distinguish themselves with different conditions.


For notational convenience, we define the signal-to-noise ratio (SNR), interference-to-noise ratio (INR), and signal-to-interference ratio (SIR) as:
\begin{IEEEeqnarray}{C}\label{snr}
    \text{SNR}_i = \frac{|h_{ii}|^2 p_i}{\sigma^2},
    \text{INR}_{ji} = \frac{|h_{ji}|^2 p_j}{\sigma^2}, 
    \text{SIR}_{ji} = \frac{|h_{ii}|^2 p_i}{|h_{ji}|^2 p_j}.
\end{IEEEeqnarray}  
  
\textbf{FlashLinQ.}
The link selection criterion of FlashLinQ\cite{Flash} is that, a link $D_i$ can be selected if it does not experience significant interference from and does not cause significant interference to other links in the selected set $\S$, i.e.,
\begin{IEEEeqnarray}{rCl}\label{eq_flash1}
\text{FlashLinQ: } \left \{ \begin{matrix}
\text{SIR}_{ij} \geq \theta, \; \forall j \in \S, \\
\frac{|h_{ii}|^2 p_i}{\sum_{j \in \S} |h_{ji}|^2 p_j} \geq \theta,
\end{matrix}
\right.
\end{IEEEeqnarray}
where $\theta$ is a predefined threshold. The effectiveness of FlashLinQ significantly hinges on the parameter $\theta$, and selecting an appropriate $\theta$ can pose practical challenges.

\textbf{ITLinQ \& ITLinQ+.}
Both ITLinQ \cite{IT} and ITLinQ+ \cite{IT+} are inspired by the TIN optimality conditions, which are relaxed and translated into two criteria that can be locally assessed in a distributed manner:
\begin{IEEEeqnarray}{rCl}\label{eq_IT1}
\text{ITLinQ: } \left \{ \begin{matrix}
M\text{SNR}^\eta _{i} \geq \max_{j \in \S} \text{INR}_{ji}, \\
M\text{SNR}^\eta _{i} \geq \max_{j \in \S} \text{INR}_{ij},
\end{matrix}
\right.
\end{IEEEeqnarray}
and
\begin{IEEEeqnarray}{rCl}\label{eq_IT+1}
\text{ITLinQ+: } \left \{ \begin{matrix}
\text{SNR}^\eta_{i} \geq \max_{j \in \S} \left\{
    \frac{\text{INR}_{ji}}{(\min_{k \in \S, k \neq j} \text{INR}_{jk})^\gamma}
\right\},
\\

\text{SNR}^\eta_{i} \geq \max_{j \in \S} \left\{
    \frac{\text{INR}_{ij}}{(\min_{k \in \S, k \neq j} \text{INR}_{kj})^\gamma}
\right\},
\end{matrix}
\right.
\end{IEEEeqnarray}
where $M$, $\eta$ and $\gamma$ are designing parameters. Similar to FlashLinQ,  the performance of ITLinQ and ITLinQ+ is highly sensitive to the values of the design parameters. Nevertheless, it is difficult to choose the best parameters in practice.

\textbf{FPLinQ.}
In FPLinQ\cite{FPLinQ}, the original combinatorial problem in Eq-(\ref{obj_func}) is reformulated as an equivalent fractional programming form, by which a sub-optimal solution is proposed to iteratively solve the reformulated problem within a finite number of iterations. Each iteration analytically determines the link schedules with the help of auxiliary variables, thus creating a centralized mechanism for scheduling. FPLinQ is considered the state-of-the-art model-based link scheduling method in terms of data rate performance, although it incurs high computational complexity due to the iterative algorithm and the centralized computation. Its ability to achieve superior performance in comparison to other methods makes it highly valued in the development of data-driven approaches by providing high-quality solved instances of training datasets. 
\subsubsection{Data-driven approaches}
In contrast to the above model-based methods, data-driven approaches learn a complex mapping/function, e.g., a link scheduler, from training data. In particular, SpatialLinQ and GELinQ are two typical data-driven link scheduling approaches, learning an end-to-end link scheduler either through unsupervised or supervised learning by extracting certain patterns from solved link selection instances generated from e.g., FPLinQ.

\textbf{SpatialLinQ.}
SpatialLinQ\cite{spatial} relaxes the need for CSI to the geographic locations of transmitters/receivers for link scheduling. By employing convolutional neural networks to learn the geographical locations of interfering or interfered nodes, SpatialLinQ determines if certain links are selected. 
While competitively effective compared to FPLinQ without requiring precise CSI, a notable challenge is a substantial demand for a large amount of training samples, posing constraints on both memory and time resources during training.

\textbf{GELinQ.}
By representing the D2D network as an interference graph, a fully-connected directed graph, GELinQ\cite{GNNLinQ} utilizes graph embedding to convert the D2D network into a low-dimensional space. Then it addresses the binary classification problem with a multi-layer classifier. Despite reducing the required number of training samples, this approach is trained with solved instances being labels generated by FPLinQ, experiencing some compromise in performance.

It is worth noting that both supervised and unsupervised training methods have been explored by SpatialLinQ\cite{spatial} and GELinQ\cite{GNNLinQ}. As the supervised versions employ datasets generated by FPLinQ for training, their sum rates are inherently limited by the best results of FPLinQ. While the unsupervised approach exhibits a marginal improvement in sum rates compared to the supervised counterpart, there is still some gap to FPLinQ. These data-driven approaches hinge critically on training data and heuristics, leaving aside wisdom/insights obtained from model-based methods. 

\subsection{Power Control}
State-of-the-art power control mechanisms can be classified into model-based approaches, such as WMMSE \cite{WMMSE} and FPLinQ-pc \cite{FPLinQ}, model-driven, such as UWMMSE \cite{UWMMSE}, and data-driven approaches, such as PCGNN \cite{PCGNN}, among others.

\subsubsection{Model-based approaches}
WMMSE \cite{WMMSE} and FPLinQ-pc \cite{FPLinQ} both solve power control problems using mathematical optimization techniques. WMMSE leverages the block coordinate descent method, iteratively minimizing a reformulated weighted mean-squared-error cost function. FPLinQ-pc, a version of FPLinQ specifically for power control, addresses the issue from the fractional programming perspective, sharing a similar procedure with the original FPLinQ approach. 

\subsubsection{Model-driven approaches}
Model-driven deep learning techniques integrate iterative communication algorithms with deep learning tricks, thereby reducing computational resources and training duration \cite{he2019model}. For example, UWMMSE \cite{UWMMSE} ``unfolded'' the iterative WMMSE algorithm into a layered structure, where each layer mimics one iteration with certain adjustable parameters
optimized through learning. Compared to the classical iterative WMMSE, UWMMSE leverages the learning-to-optimize mechanisms with GNNs to achieve faster convergence and comparable optimization performance.

\subsubsection{Data-driven approach}
PCGNN \cite{PCGNN} leverages the message passing GNNs to solve the graph optimization problem of power allocation in an end-to-end data-driven manner. Each GNN layer consists of the aggregation and combination operations parameterized by multi-layer perceptrons (MLPs) to produce updated hidden representations from information obtained from the neighboring nodes.
PCGNN is terminated after a number of GNN layers with an MLP layer to map hidden representations to power allocation variables.

Notably, these power control mechanisms, no matter model-based or model/data-driven, all require exact CSI, and their computational complexity is $O(N^2)$.

\section{Hybrid Model/data-driven Graph Reinforcement Learning Spectrum Sharing}\label{Model/data-driven Graph Reinforcement Learning Link Scheduling}

\subsection{Overview of Our Approach}\label{Overview}
As mentioned above, the model-based link scheduling and all previously reviewed power control approaches rely critically on the accuracy of CSI.
The data-driven approaches require a large number of samples or labeled training samples generated from the solved instances by e.g., FPLinQ. 


To relax the requirements of accurate CSI and massive training samples, we propose a hybrid model/data-driven framework that leverages the advantages of both model-based and data-driven methods. Our framework incorporates a specially designed Markov decision process (MDP) tailored for each task. We designate the approach for link scheduling as GRLinQ and for power control as GRLinQ-pc. Both GRLinQ and GRLinQ-pc models share the same network architectures and feature designs, yet differ in their MDP components, allowing each to effectively tackle its specific interest.

Motivated by \cite{khalil2017learning,lwd,shan2023learning,Liang2019,Liang2020}, 
we employ a graph reinforcement learning (GRL) framework to (1) disseminate the decision-making process across multiple iterations via reinforcement learning (RL); and (2) parameterize both the policy and value networks with graph neural networks (GNNs).
The use of RL relaxes the need for obtaining solved instances as training data, where the network layouts without link scheduling solutions are sufficient.
By translating some domain experts' knowledge into node features, GRLinQ achieves significant improvements over the state-of-the-art benchmarks. 

\begin{figure*}[tbp]
\begin{center}
     \includegraphics[width=0.9\textwidth]{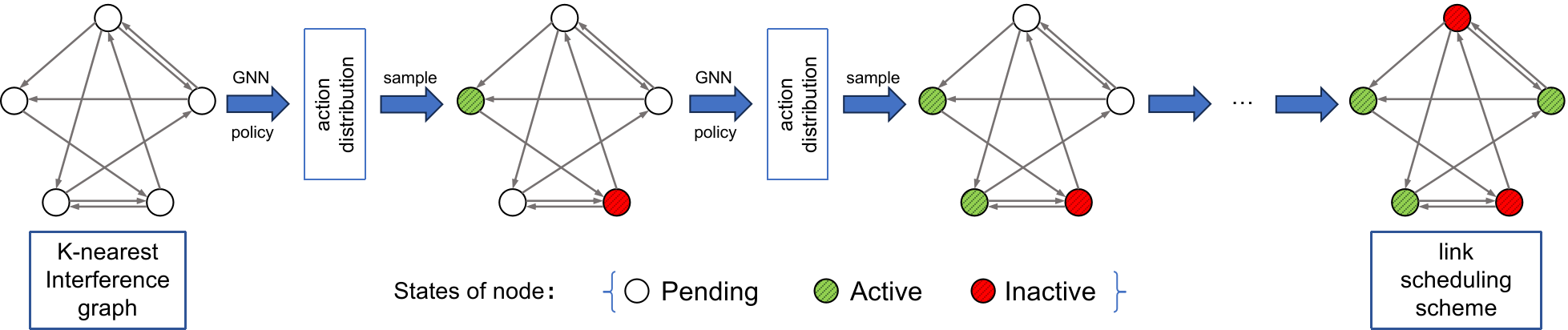}
\end{center}
   \caption{Illustration of the proposed framework. All D2D pairs are initially set to a pending state. During each iteration, the policy network makes decisions based on the current state. The policy network has considerable flexibility to classify any number of D2D pairs as active, inactive, or to retain them in the pending state. The model terminates once all D2D pairs have exited the pending state.}
   \label{framework}
\end{figure*}

Our approach takes the $K$-nearest interference graph $\gG_d = (\gV, \gE)$ as the input, where $\gV$ represents the set of nodes and $\gE$ represents the set of edges. When generating the $K$-nearest interference graph, we consider the Tx-Rx distance instead of the channel gain, hence eliminating the need for CSI. The node features are specified later. At each iteration, the agent (policy network) makes a decision to some of vertices and postpones the remaining vertices to later iterations. This process will be repeated until all vertices have been given decisions. The framework of GRLinQ is illustrated in Figure \ref{framework}. GRLinQ-pc follows the same framework as GRLinQ, but replaces the discrete states and actions with continuous ones, as will be detailed below. Thanks to the inherent randomness of RL, the same test can be conducted in parallel without loss of robustness and reliability.

\begin{figure*}[t]
\begin{center}
   {%
     \includegraphics[width=0.9\textwidth]{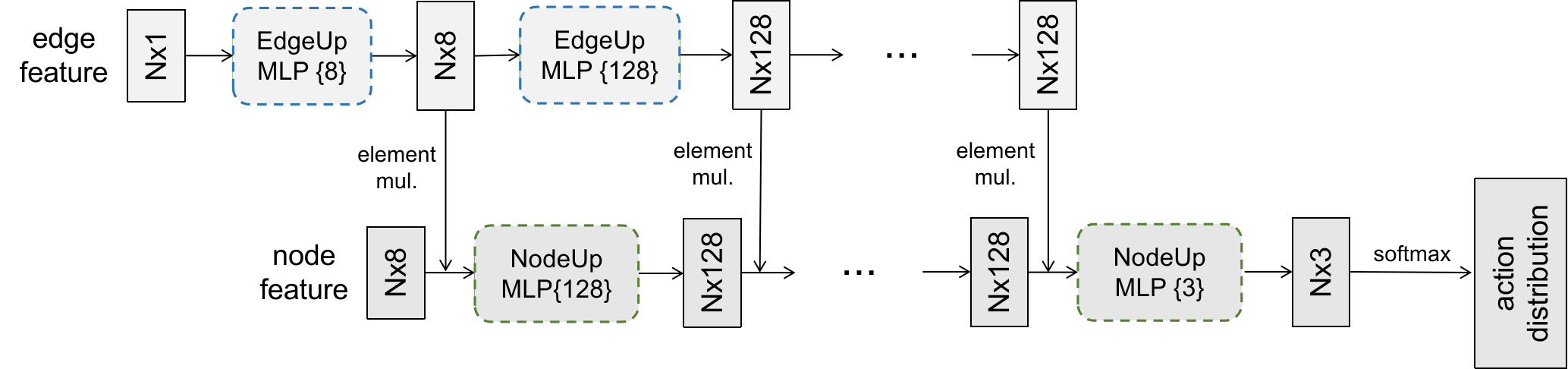}}
\end{center}
\caption{
\textbf{Policy network architectures:} The architecture of our proposed MPGNN policy network. Edge features are first updated by an EdgeUpdate block. Following the edge update, node features are combined with edge features and then updated by the NodeUpdate block. These processes repeat for $L$ layers, and the final layer produces action distributions using the updated node features.\\
\textbf{EdgeUp block:} This block is a multi-layer perceptron (MLP) that updates the edge features. This transformation is applied at both the initial stage and during subsequent layers.\\
\textbf{NodeUp block:}  This block aggregates messages from neighboring nodes and then applies an MLP to the concatenated node features and aggregated messages, resulting in updated node features.
}
\label{GNN_structure}
\end{figure*}

\subsection{Reinforcement Learning}
In what follows, we describe the distinct RL procedures separately for link scheduling and power control.
\subsubsection{Link scheduling}
We represent the RL procedure as Markov decision processes (MDPs) with four components:

\textbf{State.}
A state $\vs = [s_i:i \in \gV]$ is a sequence of operations on the given graph $\gG_d$. Each node's state can be one of three possibilities: $\{active,inactive,pending\}$. Initially, the state $\vs$ is set such that all elements are in the ``pending'' state. The algorithm terminates either when no nodes in the graph remain in the ``pending'' state or when the iterations limit, denoted as $T$, is reached.

\textbf{Action.}
Given a state, the agent will produce an action $\va = [a_i:i \in \gV]$. Each node's action also assumes one of the three previously mentioned states. This signifies that the agent has the flexibility to make decisions regarding ``pending'' nodes—either scheduling them as ``active'' or ``inactive''—or to defer the decision by maintaining them in the ``pending'' state for future iterations.

\textbf{Transition.}
The RL transits from state $\vs$ to the next state $\vs'$ through overwriting (or updating) the previous state with the action $\va$.

\textbf{Reward.}
The reward function, denoted as $r(\vs,\va)$, is defined as the difference in the objective function Eq-(\ref{obj_func}) before and after applying action $\va$ and transitioning to a new state $\vs'$. An additional penalty term of $-\alpha$ $( \alpha > 0)$ is added into the reward for each iteration to prompt the model to make decisions more expeditiously. Formally,
\begin{IEEEeqnarray}{rCl}\label{eq_reward}
r(\vs,\va) =  \sum^N_{i=1}w_i R_i(\vx') - \sum^N_{i=1}w_i R_i(\vx) - \alpha,
\end{IEEEeqnarray}
where $\vx$ and $\vx'$ are mapped from $\vs$ and $\vs'$ separately. $x_i = 1$ if $s_i$ is \text{``active''}, and $x_i = 0$ otherwise. The same applies to $\vx'$ and $\vs'$. We set $\alpha = 1/T$, ensuring that the model receives a sum of the penalties no bigger than 1.

\subsubsection{Power control}
The MDP of GRLinQ-pc transforms the state of each node into a continuous number between 0 and 1 to control the transmission power. The actions are designed as discrete numbers of varying scales between -0.1 and 0.1, allowing for fine-grained adjustments to the power levels. Formally, we set the MDP components as:

\textbf{State.}
A state $\vs = [s_i : i \in \gV]$ is a sequence of continuous numbers, where each element belongs to $[0, 1]$, i.e., $ s_i \in [0, 1], \forall i \in \gV.$ Initially, the state is set to $s_i = 0.5, \forall i \in \gV.$ The algorithm terminates either when the iterations limit $T$ is reached or an early stopping criterion is satisfied.

\textbf{Action.}
Given a state, the agent will produce an action $\va = [a_i:i \in \gV],$ where $$a_i\in \{-0.1,-0.01,-0.001,0,0.001,0.01,0.1\}  \quad \forall i \in \gV.$$  

\textbf{Transition.}
The RL process transitions from state $\vs$ to the next state $\vs'$ by adding the action $\va$ to the current state $\vs$. If the resulting state value exceeds the [0,1] range, the model clips it to ensure it remains within bounds.

\textbf{Reward.}
The reward function is the same as before in Eq.~(\ref{eq_reward}), where $\vx$ and $\vx'$ correspond to $\vs$ and $\vs'$ respectively.

The policy network aims to maximize the expected cumulative reward. To optimize the parameters of the policy network, we utilize proximal policy optimization (PPO) \cite{ppo}.

\subsection{GNN Policy and Value Network Architecture}
To capture graph structures of the $K$-nearest interference graph, we employ a message passing GNN for policy and value networks, where the node features are well-crafted with experts' knowledge from model-based methods. This gives us a hybrid model/data-driven GNN architecture.
\subsubsection{Data-driven message passing GNN (MPGNN)}
There exist numerous MPGNNs specifically designed to leverage edge features, such as \cite{cui2020edge,isufi2021edgenets,monninger2023scene}. These models typically employ specialized message passing mechanisms to effectively capture edge information within the graph, thereby enhancing the model's expressive capacity and predictive accuracy. Our proposed model first updates the edge features through a neural network, then propagates the message computed by 
a combination operation
of edge and node features, and finally updates the node features through a neural network. The policy network architecture is showcased in Figure \ref{GNN_structure}. At the $l$-th layer, our proposed MPGNN is structured as follows:

\textbf{Edge Update:} The edge feature is first updated by a multi-layer perceptron (MLP) $\vf_e^{(l)}$:
\begin{IEEEeqnarray}{rCl}
\ve_{ji}^{(l)} = \vf_e^{(l)}(\ve_{ji}^{(l-1)}), \label{Eq_edge_update}
\end{IEEEeqnarray}
where $\ve_{ji}^{(l-1)}$ represents the edge feature of edge $(j,i)$ at the $(l-1)$-th layer.

\textbf{Node Update:}
Each node then sends a message to its outgoing neighbors.
The message sent from $v_j$ to $v_i$ is computed by a message function $\mathrm{msg}^{(l)}$:
\begin{IEEEeqnarray}{rCl}
\vm_{ji}^{(l)} = \mathrm{msg}^{(l)}(\ve_{ji}^{(l)},\vv_j^{(l-1)}),
\end{IEEEeqnarray}
where $\vv_j^{(l-1)}$ is the feature vector of node $j$ at the $(l-1)$-th layer. For simplicity, an element-wise multiplication is applied for the $\mathrm{msg}^{(l)}(\cdot)$ function. 
Each node $i$ aggregates the messages sent from its incoming neighbors $\mathcal{N}(i)$ with a summation:
\begin{IEEEeqnarray}{rCl}\label{message passing equation}
\va_i^{(l)} &=& \sum_{j\in \mathcal{N}(i)}  \vm_{ji}^{(l)}.
\end{IEEEeqnarray}


Afterwards, each node updates its own feature through an MLP applied to the concatenation of its own feature and the aggregated messages, i.e., 
\begin{IEEEeqnarray}{rCl}
\label{message passing equation2}
\vv_i^{(l)} &=& \vf_n^{(l)}(\text{concat}(\vv_i^{(l-1)} | \va_i^{(l)})),
\end{IEEEeqnarray}
where $\vf_n^{(l)}$ is an MLP with the \textit{ReLU} activation function at the $l$-th layer. 

To generate actions and value estimations in the final layer, the policy and value networks employ softmax and graph read-out functions with sum pooling \cite{gin} instead of ReLU.
\subsubsection{Model-driven feature design}\label{Feature transformation}


Drawing from expert knowledge, we formulate three types of node features inspired by FlashLinQ, ITLinQ, and ITLinQ+. 

First, we frame the selection criteria into node features, relating to SNR, INR, and SIR. To provide greater flexibility to the GNN, we eliminate manually designed variables as much as possible. For FlashLinQ, we take the logarithm of the left-hand side of Eq-(\ref{eq_flash1}) as the features. As for ITLinQ and ITLinQ+, we first take the logarithm of both sides of Eq-(\ref{eq_IT1}) and Eq-(\ref{eq_IT+1}), exclude the designing parameters, and treat both sides of the resulting expressions as features separately. It is worth noting that in order to maintain the right-hand sides of Eq-(\ref{eq_IT+1}) as an entity and avoid creating excessive initial features, we set $\gamma = 0.1$, consistent with \cite{IT+}. 

Second, instead of considering channel gains $|h_{ij}|^2$, we consider the pairwise distance $d_{ij}$ between transmitters and receivers as node features, as $|h_{ji}|^2$ is positively correlated with $\frac{1}{d_{ji}}$.
As such, we replace $|h_{ii}|^2$ and $|h_{ji}|^2$ in node features with $\frac{1}{d_{ii}}$ and $\frac{1}{d_{ji}}$, respectively, where 
$d_{ji}$ represents the distance between $\text{Tx}_j$ and $\text{Rx}_i$.  For simplicity, we keep $p_i$ and $\sigma$ fixed and exclude them from our feature set. 

In addition, we include the current node state of the MDP, $s_i$, and the current iteration count, $t$, to the feature sets. The state $s_i$ is mapped to a one-hot vector, and the current iteration $t$ is normalized by the maximum number of iterations $T$. As a result, the final sets of initial node features are shown in Table \ref{table_feature}. We have also included the model with no experts' knowledge (a.k.a. pure-data) as a baseline. 
Finally, the initial feature of edge $(j,i)$ is set as $\log(\frac{1}{d_{ji}})$ for all models. The D2D distances in each network layout are normalized by the maximum distance for consistency. 

\subsection{Computational Complexity}
The complexity of GRLinQ arises from the feature generation and GNN message passing, i.e., Eq-(\ref{message passing equation}) and Eq-(\ref{message passing equation2}).
Recall that $N, L, K$ represent the numbers of D2D links, MPGNN layers, and the maximum MDP iterations, respectively.
For GRLinQ with ITLinQ+ features, the complexity of feature generation 
is $O(K^2)$, since the $K$-nearest interference graph, instead of a fully connected graph, is adopted for MPGNN. 
The complexity of the entire feature generation process, where every node computes features in each MDP iteration, is given by
$O(N T K^2)$,
which simplifies to $O(N)$ since $K$ and $T$ are constants. Similarly, the complexity of the message passing process can be computed as
$O(N T K L)$,
which also simplifies to $O(N)$ as $L$ is a constant as well. Therefore, the overall complexity of GRLinQ is $O(N)$.

\begin{table}[htbp]\centering
\vspace{-5pt}
\caption{Model-driven node feature designs}
\label{table_feature}
\resizebox{1\columnwidth}{!}{
\begin{tabular}{|c|c|}
\hline
\makecell{Inspiration} & Node Features  \\ \hline\hline
\makecell{pure-data} & $s_i,t,d_{ii}$  \\ \cline{1-2}
FlashLinQ & \makecell{$s_i,t,\log(d_{ii}),\min_{j \in \S} \{\log(d_{ij})-\log(d_{jj})\}$,\\ $\log(\frac{1}{d_{ii}})-\log(\sum_{j \in \S}\tfrac{1}{d_{ji}})$} \\ \cline{1-2}
ITLinQ & \makecell{$s_i,t,\log(d_{ii}),$\\$ \max_{j \in \S} \log(\frac{1}{d_{ji}}),\max_{j \in \S} \log( \frac{1}{d_{ij}})$}  \\ \cline{1-2}
ITLinQ+ & \makecell{$s_i,t,\log(d_{ii}),$\\ $\max_{j \in \S}\{ \log(\frac{1}{d_{ji}})-\gamma  \min_{k \in \S \setminus  j} \log(\frac{1}{d_{jk}})\},$\\$\max_{j \in \S}\{ \log(\frac{1}{d_{ij}})-\gamma \min_{k \in \S \setminus j} \log(\frac{1}{d_{kj}})\}$ }\\ \hline
\end{tabular}
}
\end{table}

\begin{table}[h]
\centering
\vspace{-10pt}
\caption{System Parameters}
\label{Table_system_para}
\resizebox{0.75\columnwidth}{!}{
\begin{tabular}{|c|c|}
\hline
Parameter & Value \\ \hline \hline
Area size & 500 m $\times$ 500 m \\ \hline
Tx-Rx distance & 2 m $\sim$ 65 m \\ \hline
Noise spectral density & -169 dBm/Hz \\ \hline
Bandwidth & 5 MHz \\ \hline
Carrier frequency & 2.4 GHz \\ \hline
Antenna height & 1.5 m \\ \hline
Maximum transmit power & 40 dBm \\ \hline
\end{tabular}
}
\end{table}

\section{Experiments}\label{Experiments}
In this section, we evaluate the performance and generalizability of our proposed GRLinQ and GRLinQ-pc. In the experiments, we set
$w_i = 1$ for all $i$ for simplicity.

\textbf{Network setting.}
To facilitate comparison, we employ a setup consistent with that of \cite{spatial, GNNLinQ}, as summarized in Table \ref{Table_system_para}. Specifically, we consider $N$ transmitters uniformly positioned within a 500 m $\times$ 500 m square area. The receivers are uniformly distributed within a range of 2 to 65 meters from their paired transmitters. The transmitter-receiver channels follow the short-range outdoor model ITU-1411 with a distance-dependent path-loss \cite{ITU-R-P1411-8}. 
Unless otherwise specified, the carrier frequency is set to 2.4 GHz, and the antenna height is 1.5 m. The transmit power and noise spectral density are set to 40 dBm and -169 dBm/Hz, respectively.



\textbf{GRLinQ settings.}
For the $K$-nearest interference graph adopted for MPGNN, we choose $K=10$ for scalability, whilst a higher $K$ does not make a significant difference, as discussed later in Section \ref{Sec_Scalability}. Unless otherwise specified, GRLinQ and GRLinQ-pc are equipped with node features inspired by ITLinQ+. The impact of different feature designs is detailed in Section \ref{Sec_feature_design}. We use a trivial map instead of an MLP in the edge update process, as discussed in Section \ref{Sec_Ablation}.

We set the number of MPGNN layers $L=4$ and the hidden dimension $h = 128$.
%
We set the RL iteration limit $T=32$, with an early stopping strategy. For each network setting, we generated 1,000 network layouts as a testing set. Each layout represents a topology graph with randomly distributed transmitters and receivers.
Model-based methods are directly applied to the test set. Learning-based methods are trained on different numbers of layouts, according to their requirements, and subsequently evaluated on the test set. The proposed GRLinQ is trained with 500 network layouts. We report the average sum rate ratio, which denotes the ratio of the sum rate achieved by the method to that achieved by FPLinQ, averaged over the 1,000 test layouts. 

\subsection{Comparison with State-of-The-Art Methods}
We compare the state-of-the-art link scheduling or/and power control approaches in three aspects: 
\begin{itemize}
    \item For link scheduling, we compare GRLinQ with other link scheduling approaches.
    \item For power control, we compare GRLinQ-pc with other power control approaches.
    \item Under the same network settings, we jointly compare link scheduling and power control approaches.
\end{itemize}


\begin{table}[tbp]
\vspace{-0pt}
\centering
\caption{Average sum rate ratios achieved by link scheduling and power control approaches (higher is better)}
\label{Table_joint}
\resizebox{0.9\columnwidth}{!}{
\begin{tabular}{|c|c|c|}
\hline 
Method &CSI& Average Sum Rate Ratio \\ \hline
All scheduled &No& 0.656 \\ \hline
FlashLinQ \cite{Flash}&Yes& 0.776 \\ \hline
ITLinQ \cite{IT}&Yes& 0.840 \\ \hline
ITLinQ+ \cite{IT+}&Yes& 0.877 \\ \hline
GELinQ \cite{GNNLinQ}&No& 0.952 \\ \hline
Greedy &Yes& 0.971 \\ \hline
SpatialLinQ \cite{spatial}&No& 0.984 \\ \hline
FPLinQ \cite{FPLinQ}&Yes& 1.000 \\ \hline
\textbf{GRLinQ} &No& 1.012 \\ \hline \cline{1-3}
UWMMSE-4* \cite{UWMMSE}&Yes& 1.034 \\ \hline
PCGNN-1* \cite{PCGNN}&Yes& 1.047 \\ \hline
PCGNN-2* \cite{PCGNN}&Yes& 1.049 \\ \hline
UWMMSE-8* \cite{UWMMSE}&Yes& 1.054 \\ \hline
\textbf{GRLinQ-pc*} &No& 1.055 \\ \hline
FPLinQ-pc* \cite{FPLinQ}&Yes& 1.059 \\ \hline
WMMSE* \cite{WMMSE}&Yes& 1.060 \\ \hline
\end{tabular}
}
\end{table}

\textbf{1) Link scheduling.}
We first compare our proposed GRLinQ against the state-of-the-art link scheduling benchmarks:
\begin{itemize}
    \item \textbf{All scheduled.} Activating all the links.
    \item \textbf{FlashLinQ.} Run default FlashLinQ \cite{Flash}, using the selection criterion in  Eq-(\ref{eq_flash1}).
    \item \textbf{ITLinQ.} Run default ITLinQ \cite{IT}, using the selection criterion in  Eq-(\ref{eq_IT1}).
    \item \textbf{ITLinQ+.} Run default ITLinQ+ \cite{IT+}, using the selection criterion in  Eq-(\ref{eq_IT+1}).
    \item \textbf{GELinQ.} Run default GELinQ \cite{GNNLinQ}, trained with 500 layouts solved by FPLinQ.
    \item \textbf{Greedy.} Sort links based on distance and activate one by one if increasing the overall sum rate.
    \item \textbf{SpatialLinQ.} Run default SpatialLinQ \cite{spatial}, trained with 800,000 layouts.
    \item \textbf{FPLinQ.} Run FPLinQ \cite{FPLinQ} with all ``ones'' initialization of $\vx$ and 100 iterations.
\end{itemize}

We set the number of D2D links to be 50. 
As shown in the upper section of Table \ref{Table_joint}, our proposed method, GRLinQ, even goes beyond the optimization-based approach FPLinQ by a margin of 1.2\%, with a substantial reduction in CSI requirement.
Moreover, compared with the state-of-the-art data-driven methods, SpatialLinQ and GELinQ, our proposed GRLinQ achieves almost 3\% and 6\% improvements, respectively, with the same number of training samples as GELinQ.

\begin{figure}[t]
\begin{center}
     \includegraphics[width=0.95\columnwidth]{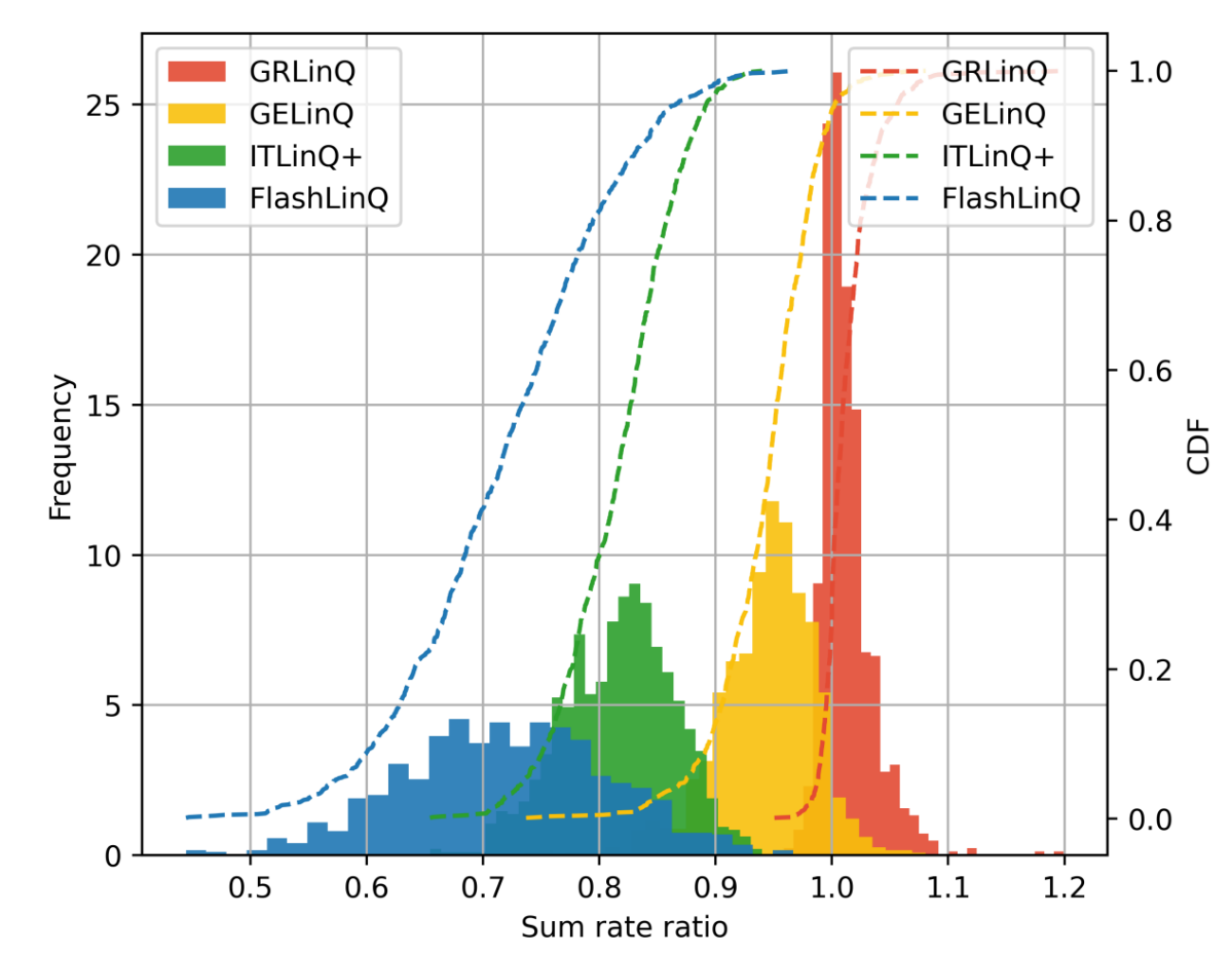}
\end{center}
 \vspace{-15pt}
   \caption{CDF of sum rate ratios for GRLinQ, GELinQ, ITLinQ+, and FlashLinQ. The number of D2D links is set to be 50.}
   \label{Figure_Distribution}
\end{figure}

Figure \ref{Figure_Distribution} illustrates the cumulative distribution function (CDF) of sum rate ratio for GRLinQ, GELinQ\footnote{The results of GELinQ were obtained by running the code provided by \cite{GNNLinQ} at https://github.com/ZhiweiShan/graph\_embedding\_link\_scheduling.}, ITLinQ+ and FlashLinQ against FPLinQ. It can be seen that GRLinQ demonstrates a substantial gain over others. Remarkably, even without utilizing CSI during online testing, GRLinQ strictly outperforms FPLinQ in the majority (722 out of 1000) of samples. This observation implies that, 
GRLinQ could potentially discover more advantageous link scheduling policies than the best-known optimization-based methods, e.g., FPLinQ.


\textbf{2) Power control.}
We then compare the proposed GRLinQ-pc with state-of-the-art power control benchmarks: 
\begin{itemize}
    \item \textbf{UWMMSE-4.} UWMMSE \cite{UWMMSE} with 4 unfolding layers, as suggested in \cite{UWMMSE}.
    \item \textbf{UWMMSE-8.} UWMMSE \cite{UWMMSE} with 8 unfolding layers.
    \item \textbf{PCGNN-1.} PCGNN \cite{PCGNN} trained with 5,000 layouts.
    \item \textbf{PCGNN-2.} PCGNN \cite{PCGNN} trained with 50,000 layouts.
    \item \textbf{FPLinQ-pc.} Run FPLinQ-pc \cite{FPLinQ} with all ``ones'' initialization of $\vx$ and 100 iterations.
    \item \textbf{WMMSE.} Run WMMSE \cite{WMMSE} with 100 iterations.
\end{itemize}

For UWMMSE, in addition to the default settings (4 unfolding layers as suggested in \cite{UWMMSE}), we also increased the number of unfolding layers to 8 to provide improved solutions. We use 5,000 training layouts for UWMMSE, as more layouts did not bring any improvement. For PCGNN, we report results for models trained with 5,000 and 50,000 layouts. 

Power control can be viewed as a continuous version of link scheduling. Therefore, we aim to compare these methods together to provide a more comprehensive understanding of their relative performance. To provide a clear view, we have marked power control mechanisms with an asterisk (*) in Table \ref{Table_joint}. It can be immediately observed that, under this network setting with 50 D2D links, all power control methods outperform the link scheduling methods. GRLinQ-pc achieved remarkable performance using just 500 training layouts, outperforming other learning-based methods UWMMSE and PCGNN. It performs only about 0.5\% worse compared to the optimization-based methods FPLinQ-pc and WMMSE, without explicitly knowing the channels and having lower computational complexity. On the other hand, GRLinQ stands out as the best link scheduling method in this context, being only 2.2\% behind the power control method UWMMSE-4.

\begin{figure}[tbp]
\begin{center}
{\includegraphics[width=0.5\textwidth]{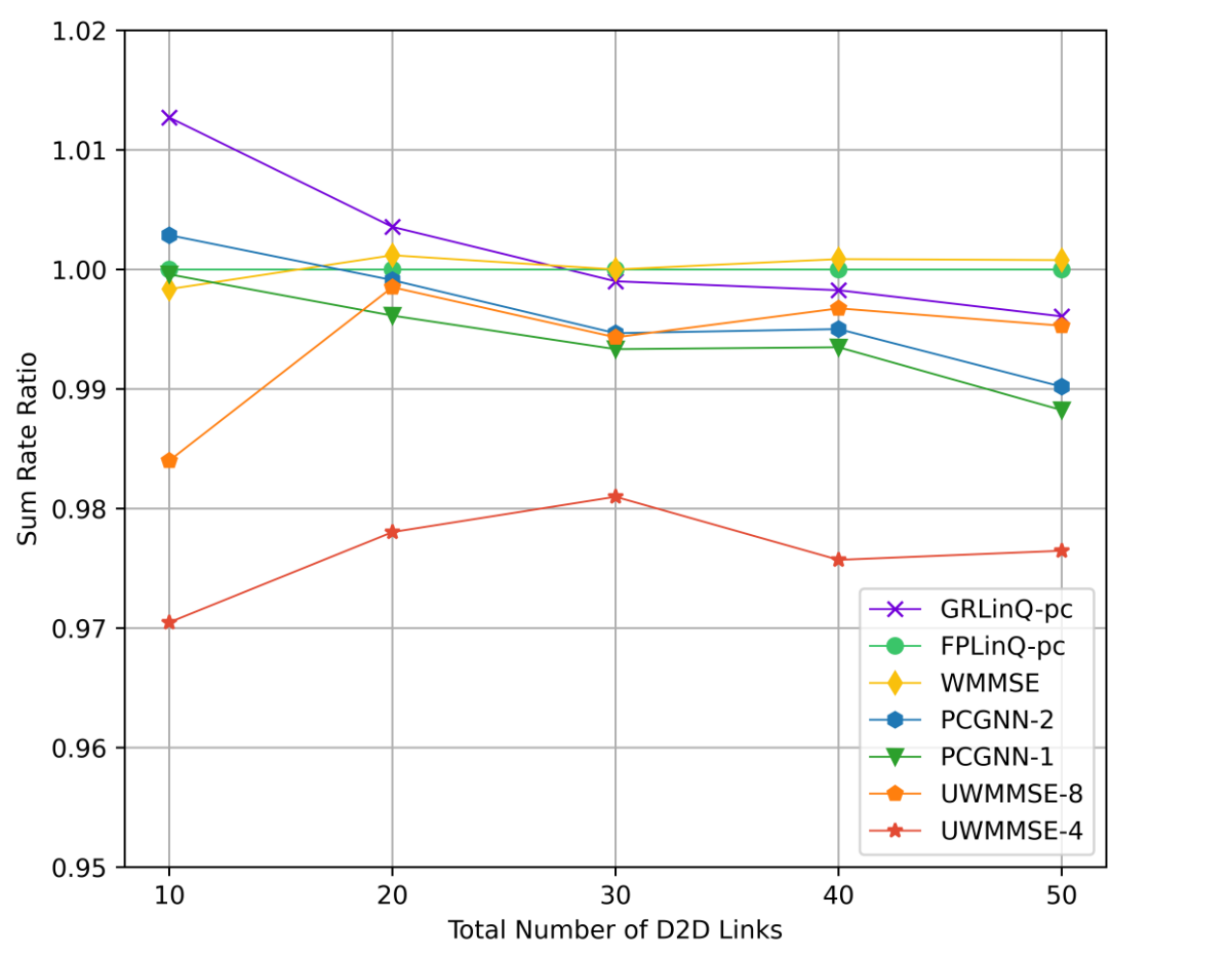}}
\end{center}
   \caption{Average sum rate ratios achieved by different power control approaches, GRLinQ-pc, FPLinQ-pc, WMMSE, PCGNN, UWMMSE.}
   \label{PC_SOTA}
\end{figure}

Since power control is particularly challenging, it is typically implemented on small networks (fewer than 30 D2D links) \cite{WMMSE,UWMMSE,PCGNN}. Therefore, we then compare power control methods on networks with 10 to 50 D2D links, as shown in Figure \ref{PC_SOTA}. It can be observed that when the network size is small, GRLinQ-pc performs the best, even surpassing WMMSE and FPLinQ-pc. As the number of links increases, GRLinQ-pc's performance slightly declines but consistently outperforms other learning-based methods with CSI.

\begin{figure}[tbp]
\begin{center}
  {%
     \includegraphics[width=0.5\textwidth]{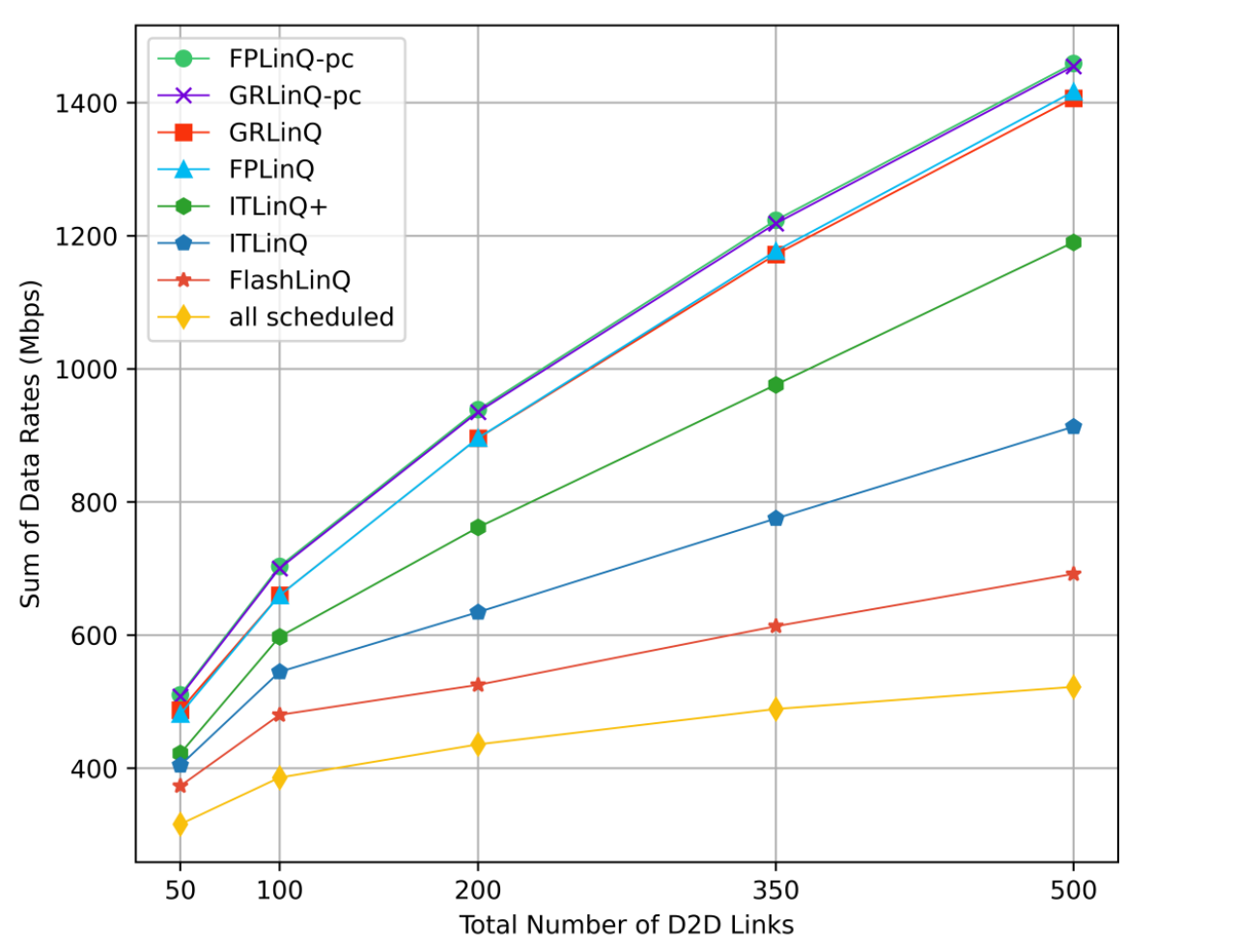}}
\end{center}
   \caption{Sum rates versus the total number of D2D links. GRLinQ and GRLinQ-pc are trained and tested under each number of links setting.}
   \label{MBPS_train}
\end{figure}

\textbf{3) Joint link scheduling and power control.}
We then jointly apply the proposed GRLinQ and GRLinQ-pc in large-scale networks with D2D link counts of 50, 100, 200, 350, and 500, using FPLinQ, FPLinQ-pc, and other link scheduling methods as benchmarks. Figure \ref{MBPS_train} illustrates the average sum rate in Mbps. GRLinQ and GRLinQ-pc demonstrate performance that is comparable to or even better than FPLinQ and FPLinQ-pc, highlighting their effectiveness in large-scale network scenarios. Specifically, GRLinQ closely matches FPLinQ, and when the number of D2D links is smaller than 200, GRLinQ slightly outperforms FPLinQ. Although GRLinQ-pc performs marginally weaker than FPLinQ-pc across all network sizes, the difference consistently remains within 0.5\%. This indicates that while GRLinQ-pc is slightly behind FPLinQ-pc in large-scale networks, it still maintains a highly competitive performance level with substantially reduced computational complexity.

\subsection{Topology Awareness and Scalability}\label{Sec_Scalability}
We use a $K$-nearest interference graph instead of a fully connected graph, significantly enhancing the scalability of the GRLinQ framework. This approach offers three main advantages:
\begin{enumerate}
\item It enhances topology awareness by ensuring that the graph structure more accurately reflects the local interactions and dependencies among nodes.
\item In a $K$-nearest interference graph, each node’s number of in-neighbors (i.e., neighboring nodes that point to it) remains constant regardless of network size. This ensures that the amount of information each node receives during the GNN message passing process is fixed, facilitating scalability to large-scale networks.
\item It reduces both message passing and computational complexity from $O(N^2)$ to $O(KN)$.
\end{enumerate}

Notably, by training GRLinQ with 50 D2D links and then testing it on networks with 10,000 links using $K=10$, we reduced computational complexity by a factor of 1,000, while achieving 97\% of FPLinQ’s performance.

In what follows, we first examine the \textbf{impact of varying $\boldsymbol{K}$} on the model's performance to demonstrate that the model is not sensitive to $K$. Subsequently, we evaluate the model's scalability by testing it on unseen \textbf{large-scale networks} with \textbf{similar network density} and \textbf{higher network density}.

\begin{table}[htbp]
\vspace{-0pt}
\centering
\caption{Performance of GRLinQ for different $K$, where the exact value of $K$ is shown in brackets}
\label{Table_different_K}
\begin{tabular}{|c|cccc|}
\hline
$N$   & $K=10$ & $K=0.1N$ & $K=0.2N$ & $K=0.3N$ \\ \hline
50  & \textbf{1.012} (10)   & 1.010 \ (5) & \textbf{1.012} (10) & \textbf{1.012} (15) \\ \hline
100 & 1.001 (10)   & 1.001 (10) & 1.005 (20) & \textbf{1.007} (30) \\ \hline
200 & 0.999 (10)   & 0.999 (20) & \textbf{1.002} (40) & 1.000 (60) \\ \hline
\end{tabular}
\end{table}

\begin{table}[htbp]
\vspace{-0pt}
\centering
\caption{Performance of GRLinQ-pc for different $K$, where the exact value of $K$ is shown in brackets}
\label{Table_different_K_pc}
\begin{tabular}{|c|cccc|}
\hline
$N$   & $K=10$ & $K=0.2N$ & $K=0.3N$ & $K=N-1$ \\ \hline
30 & \textbf{0.999} (10) & 0.993 (6)  & \textbf{0.999} (9)   & 0.991 (29)  \\ \hline
40 & \textbf{0.998} (10)  & 0.997 (8)   & 0.997 (12)  & 0.993 (39)  \\ \hline
50 & \textbf{0.996} (10)  & \textbf{0.996} (10) & 0.990 (15) & 0.989  (49)\\ \hline
\end{tabular}
\end{table}

\textbf{1) Impact of $\boldsymbol{K}$-nearest interference graph.} 
In this section, we evaluate the impact of choosing different $K$ on GRLinQ and GRLinQ-pc performance. We hypothesize that setting $K$ to be proportional to the number of D2D pairs $N$ is beneficial. Therefore, we set $K = 0.1N$, $0.2N$, and $0.3N$ respectively, and then retrain and test the model under these configurations on three network settings with $N=50,100, \text{ and } 200$.

As shown in Table \ref{Table_different_K}, increasing $K$ sometimes provides some benefits, with the most notable improvement observed when $N=100$. However, for $N=200$, increasing $K$ from $0.2N$ to $0.3N$ results in a performance drop. This could be attributed to the declining performance of GNNs when handling very dense graphs. Overall, fixing $K=10$ already achieves satisfactory results, and
maintains computational complexity at a reasonable level.
%
Table \ref{Table_different_K_pc} shows the results of GRLinQ-pc, where a similar trend is observed. The impact of varying $K$ on the model is marginal, with differences remaining below 1\%. Therefore, we choose to set $K=10$ for all of our experiments for GRLinQ and GRLinQ-pc.


\begin{table}[tbp]
\vspace{-10pt}
\centering
\caption{Scalability of GRLinQ for large network size but similar network density}
\label{Table_Generalizability_same_density}
\resizebox{0.8\columnwidth}{!}{
\begin{tabular}{|c|c|c|}
\hline
$N$ & Area (m $\times$ m) & Average Sum Rate Ratio \\ \hline
50 & $500 \times 500$ & \multicolumn{1}{r|}{1.012 \hspace*{0.2em}(trained) }\\ \hline
100 & $700 \times 700$ & 1.003 \\ \hline
200 & $1000 \times 1000$ & 0.994 \\ \hline
500 & $1500 \times 1500$ & 0.985 \\ \hline
10,000 & $7000 \times 7000$ & 0.971 \\ \hline
\end{tabular}
}
\end{table}


\begin{table}[tbp]
\vspace{-0pt}
\centering
\caption{Scalability of GRLinQ for large network size and higher network density}
\label{Table_Generalizability}
\resizebox{0.75\columnwidth}{!}{
\begin{tabular}{|c|c|c|}
\hline 
$N$ & \multicolumn{1}{c|}{Area (m $\times$ m)} & Average Sum Rate Ratio \\ \hline
50 & \multirow{4}{*}{$500 \times 500$} & \multicolumn{1}{r|}{1.012 \hspace*{0.2em}(trained)} \\ \cline{1-1} \cline{3-3}
100 &  & 1.001 \\ \cline{1-1} \cline{3-3}
200 &  & 0.969 \\ \cline{1-1} \cline{3-3}
500 &  & 0.910 \\ \hline
\end{tabular}
}
\end{table}

\textbf{2) Large-scale networks with similar network density.} 
We first test the model's scalability to unseen networks with an increasing number of D2D links, while maintaining a similar network density. This is controlled by proportionally increasing the area size. GRLinQ is equipped with ITLinQ+ features as well. We train the model on networks with 50 D2D links over a 500 m $\times$ 500 m square area, and then tested on networks with different numbers of D2D links and different area sizes, as shown in Table \ref{Table_Generalizability_same_density}. It can be observed that the performance of GRLinQ remains competitive as FPLinQ by the increase in the number of D2D links. Notably, on a test set with 10,000 D2D pairs, which is 200 times larger than the training set, GRLinQ achieved an average sum rate ratio of 0.971, completing all decisions within $T=32$ iterations. We report only the results of GRLinQ, omitting those of GRLinQ-pc, because they share the same scalability behavior.

\textbf{3) Large-scale networks with higher network density.} We then evaluate the scalability of GRLinQ to larger and denser networks. We conducted training on the networks with 50 links as well, and subsequently tested on larger size and denser networks, maintaining the area size of 500 m $\times$ 500 m. 
The results in Table \ref{Table_Generalizability} reveal that the scalability of GRLinQ to high-density networks is acceptable. However, in extremely dense networks, specifically with $N=500$, which is ten times denser compared to the training set, GRLinQ exhibits a decline in performance. This decline may be related to the dramatically increased mutual interference in such highly dense network environments, suggesting that retraining the models with a denser training set might be necessary.
%

\begin{table}[ht]
\vspace{-5pt}
\centering
\caption{Performance of GRLinQ with different model-driven node feature designs}
\label{Table_Performance}
\resizebox{\columnwidth}{!}{
\begin{tabular}{|c|cccc|}
\hline
$N$   & GRLinQ-PD & GRLinQ-FL & GRLinQ-IT & GRLinQ-IT+ \\ \hline
50  & 0.940     & 1.006    & \textbf{1.012}     & \textbf{1.012}      \\ \hline
100 & 0.974     & 1.002     & \textbf{1.003}    & 1.001      \\ \hline
200 & 0.990     & 0.997     & 0.998     & \textbf{0.999}      \\ \hline
500 & 0.984     & 0.989     & \textbf{0.995}     & 0.993      \\ \hline
\end{tabular}
}
\end{table}

\begin{table}[ht]
\vspace{-10pt}
\centering
\caption{Performance of GRLinQ-pc with different model-driven node feature designs}
\label{Table_Performance_pc}
\resizebox{\columnwidth}{!}{
\begin{tabular}{|c|cccc|}
\hline
$N$   & GRLinQ-pc-PD & GRLinQ-pc-FL & GRLinQ-pc-IT & GRLinQ-pc-IT+ \\ \hline
10 & 0.981 & 1.001 & 1.012     & \textbf{1.013}      \\ \hline
20 & 0.916 & 0.996 & 0.994 & \textbf{1.004} \\ \hline
30 & 0.900 & 0.963 & 0.996 & \textbf{0.999} \\ \hline
40 & 0.881 & 0.972 & 0.993 & \textbf{0.998} \\ \hline
\end{tabular}
}
\end{table}

\begin{table}[htbp]
\vspace{-10pt}
\centering
\caption{Generalization of GRLinQ with different node features over different network sizes}
\label{Table_Generalizability_features}
\resizebox{\columnwidth}{!}{
\begin{tabular}{|c|cccc|}
\hline 
$N$ & { GRLinQ-PD} & { GRLinQ-FL} & { GRLinQ-IT} & { GRLinQ-IT+} \\ \hline
50\hspace*{0.2em}(trained) & 0.940 & 1.006 & \textbf{1.012} & \textbf{1.012} \\ \hline
100 & 0.956 & 0.996 & 0.998 & \textbf{1.001} \\ \hline
200 & 0.941 & 0.963 & 0.964 & \textbf{0.969} \\ \hline
500 & 0.828 & 0.860 & \textbf{0.910} & \textbf{0.910} \\ \hline
\end{tabular}
}
\end{table}

\subsection{Feature Design and Generalizability }\label{Sec_feature_design}
In this section, we assess the effectiveness of \textbf{model-driven features} together with the generalizability of the model with respect to \textbf{carrier frequency} and \textbf{user distribution}.

\textbf{1) Effectiveness of model-driven features.} We first compare the performance of GRLinQ equipped with different feature designs: 
\begin{itemize}
    \item \textbf{GRLinQ-PD.} GRLinQ with pure-data-driven features.
    \item \textbf{GRLinQ-FL.} GRLinQ with FlashLinQ features.
    \item \textbf{GRLinQ-IT.} GRLinQ with ITLinQ features.
    \item \textbf{GRLinQ-IT+.} GRLinQ with ITLinQ+ features.
\end{itemize}

The models are trained and tested on the same network settings, with 50, 100, 200, and 500 D2D links. Based on the performance evaluation results presented in Table \ref{Table_Performance}, we observe several noteworthy trends and insights. For the smallest network size, $N = 50$, both GRLinQ-IT and GRLinQ-IT+ achieve the highest average sum rate ratio of 1.012, outperforming GRLinQ-PD and GRLinQ-FL, which exhibit sum rate ratios of 0.940 and 1.006, respectively. As the network size increases to $N = 100$ and $N = 200$, the performance gap between the models narrows. GRLinQ-IT and GRLinQ-IT+ slightly outperform other models. In the largest network size of $N = 500$, GRLinQ-IT achieves the highest average sum rate ratio of 0.995, with GRLinQ-IT+ following at 0.993. This indicates that the TIN-condition-inspired designs provide superior performance in smaller network settings, although the diminishing performance gap suggests that their advantages become less pronounced as the network size increases.

Overall, the results demonstrate that models inspired by TIN-condition (GRLinQ-IT and GRLinQ-IT+) consistently outperform the pure-data-driven (GRLinQ-PD) and FlashLinQ-inspired (GRLinQ-FL) models across all network sizes. This performance superiority is particularly pronounced in smaller networks but persists even as the network size increases. These findings underscore the effectiveness of ITLinQ-inspired designs in optimizing the average sum rate performance in D2D networks.
A similar result is obtained when testing the performance of GRLinQ-pc equipped with different features, as shown in Table \ref{Table_Performance_pc}. Node features inspired by the TIN-condition have been shown to significantly enhance power control as well.

Afterward, we evaluate the generalizability of GRLinQ with different features, as shown in Table \ref{Table_Generalizability_features}. GRLinQ-IT and GRLinQ-IT+ consistently achieve better generalizability compared to GRLinQ-PD and GRLinQ-FL across all tested network sizes. At \(N=100\), GRLinQ-IT+ achieves the highest sum rate ratio of 1.001, followed closely by GRLinQ-IT at 0.998. This trend continues at \(N=200\), where GRLinQ-IT+ leads with a ratio of 0.969. In larger and denser networks, GRLinQ-PD and GRLinQ-FL performance diminishes more significantly compared to GRLinQ-IT and GRLinQ-IT+. These findings demonstrate that TIN-condition-inspired models, GRLinQ-IT and GRLinQ-IT+, exhibit better generalizability and maintain higher performance across various network densities and sizes. However, in extremely dense networks, e.g., at \(N=500\), which is ten times denser than the training set, GRLinQ shows certain performance degradation. 

\textbf{2) Generalizability for different carrier frequencies.}
We then examine how changes in carrier frequency affect its performance. The model has been trained under a carrier frequency of $2.4$ GHz with $N=50$. We assess the model when the carrier frequency is adjusted to $6$ GHz and $26$ GHz, and the results are presented in Table \ref{Table_freq}. FPLinQ continues to serve as the baseline, which recalculates the scheduling scheme based on the new carrier frequency. In the scenario where the carrier frequency changes from $2.4$ GHz to $6$ GHz and $26$ GHz, our model's performance only decreases by $0.5\%$ and $1.1\%$, respectively, still outperforming FPLinQ. This implies that GRLinQ has significant potential to handle various exceptional situations without generating unreasonable solutions.

\begin{table}[htbp]
\centering
\caption{Generalization performance for different carrier frequency}
\label{Table_freq}
\begin{tabular}{|c|ccc|}
\hline
Carrier frequency (GHz) & 2.4\hspace*{0.2em}(trained) & 6 & 26\\ \hline
Average sum rate & 1.012 & 1.007 & 1.001 \\ \hline
\end{tabular}
\end{table}

\textbf{3) Generalizability for different user distribution.}
Then, we evaluate the generalization performance of the model with respect to different user distributions. The model is trained on networks with D2D distances ranging from 2$\sim$65 meters and then tested on other D2D distance scenarios, as shown in Table \ref{Table_D2D_distance}. Surprisingly, it performs better on networks with D2D distances ranging from 10$\sim$50 meters and 30$\sim$70 meters, \textit{without further training}. Remarkably, on networks where all D2D distances are 30 meters, GRLinQ achieves 94.1\% of the performance of FPLinQ. This indicates that even when all D2D pairs have identical initial information (i.e., distance), GRLinQ can still perform scheduling effectively, likely due to the well-designed model-driven features.

\begin{table}[htbp]
\vspace{-5pt}
\centering
\caption{Generalization performance for different user distribution}
\label{Table_D2D_distance}
\begin{tabular}{|c|cccc|}
\hline
D2D distance (m)& 2$\sim$65 \hspace*{0.2em}(trained) & 10$\sim$50 & 30$\sim$70 & all 30\\ \hline
Average sum rate & 1.012 & 1.073 & 1.099 & 0.941 \\ \hline
\end{tabular}
\end{table}

\subsection{Ablation Study}\label{Sec_Ablation}
In this section, we perform an ablation study to evaluate the contributions of various components within the GRLinQ framework. Specifically, we analyze the following aspects: the role of \textbf{edge feature updates}, \textbf{training convergence}, and the effects of \textbf{transformation learning}.

\textbf{1) Impact of edge feature update.} To evaluate the effectiveness of the edge feature update mechanism, we compare the following two models:

\begin{itemize}
    \item \textbf{GRLinQ (EdgeUp):} This model updates edge features according to Eq-(\ref{Eq_edge_update}).
    \item \textbf{GRLinQ:} This model uses a trivial map, replacing Eq-(\ref{Eq_edge_update}) with $\ve_{ji}^{(l)} = \ve_{ji}^{(l-1)}$, and broadcasts it to the corresponding dimensions before performing element-wise multiplication with the node information.
\end{itemize}

We report their performance in Table \ref{Table_Attention}. It can be observed that while using an MLP for edge feature updates brings some improvements, employing a trivial mapping instead of a complex MLP yields comparable performance. This phenomenon can be attributed to the sufficiency of the node features, especially when these features are derived from theoretically sound model designs.

Incorporating a complex edge update mechanism, such as an MLP, introduces additional parameters, thereby increasing the model's complexity and computational cost. In scenarios where the dataset is limited or the variation in edge features is minimal, a sophisticated edge update mechanism may induce overfitting. Conversely, a simpler mapping approach can enhance generalization capabilities. As such, unless otherwise specified, we use the model without edge feature updates in our paper. Nonetheless, the performance of these models may be enhanced by incorporating edge feature updates.

\begin{table}[htbp]
\vspace{-0pt}
\centering
\caption{Comparison of edge feature update methods}
\label{Table_Attention}
\begin{tabular}{|c|cc|}
\hline
$N$   & GRLinQ & GRLinQ (EdgeUp) \\ \hline
50  & 1.012                           & 1.013        \\ \hline
100 & 1.001                           & 1.006        \\ \hline
200 & 0.999                           & 1.001        \\ \hline
500 & 0.993                           & 0.987        \\ \hline
\end{tabular}
\end{table}

\begin{figure}[htbp]
\vspace{-5pt}
\begin{center}
  {%
     \includegraphics[width=0.38\textwidth]{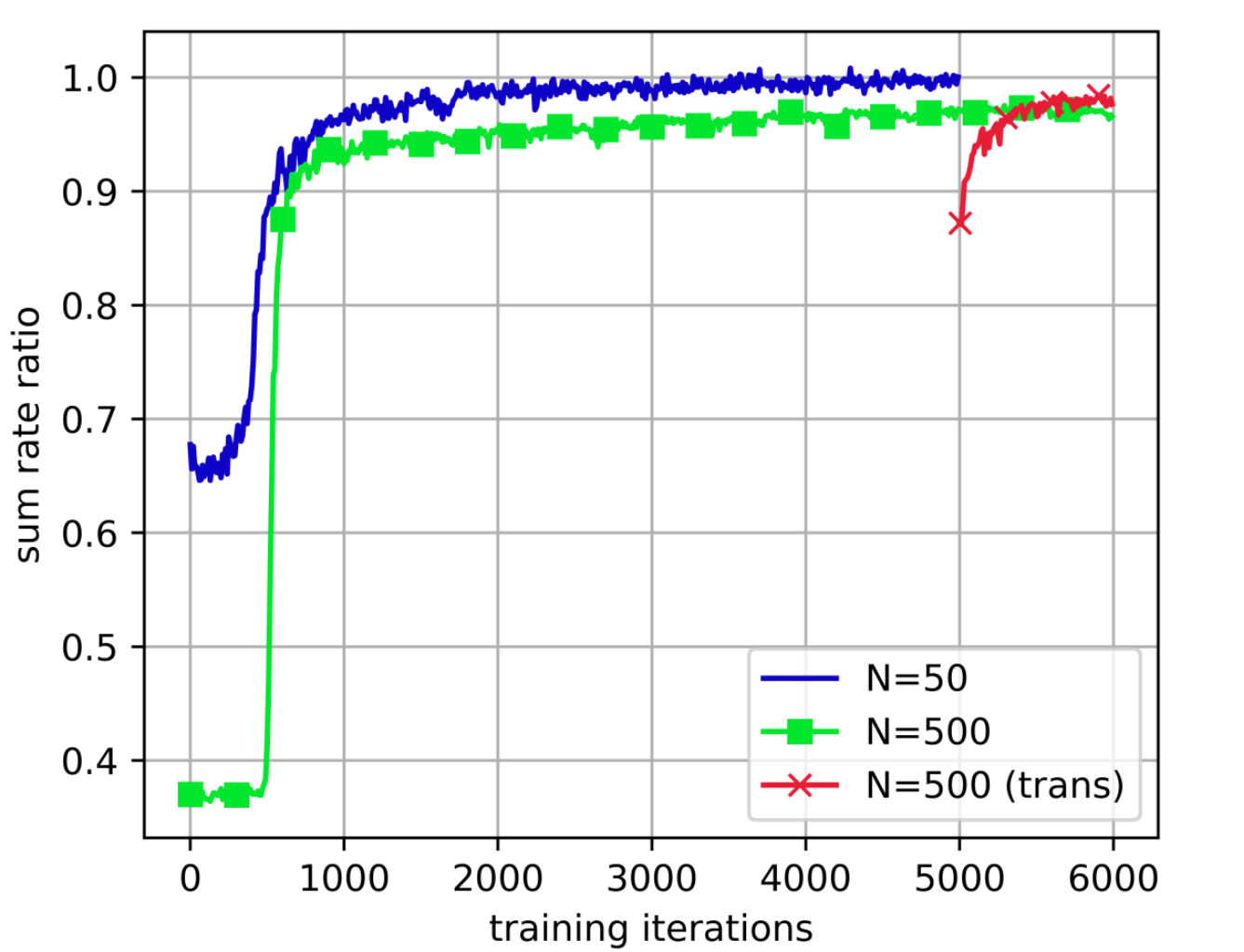}}
\end{center}
   \caption{Convergence performance of GRLinQ is evaluated under two training conditions: starting from scratch with network sizes of $N=50$ and $N=500$ separately. Additionally, the model trained with $N=50$ serves as the initial point for further training with $N=500$.
}
   \label{Convergence}
\end{figure}

\textbf{2) Training convergence and transform learning.}
In this section, we evaluate the training convergence on both small and large networks. Additionally, we employ transfer learning on large networks to accelerate the training process:

\begin{itemize}
    \item {$N=50$:} Train GRLinQ on networks with 50 D2D links for 5,000 training iterations.
    \item {$N=500$:} Train GRLinQ on networks with 500 D2D links for 6,000 training iterations.
    \item {$N=500$ (trans):} Start with the pre-trained model from $N=50$, then train for an additional 1,000 iterations on networks with 500 D2D links.
\end{itemize}

The results are shown in Figure \ref{Convergence}. For $N=50$, the model converges quickly: during the initial 400 iterations, the agent engages in extensive random exploration and experience accumulation within the environment, achieving significant performance improvement around 400 to 500 iterations. After approximately 2,000 iterations, the performance stabilizes and no longer shows significant growth.

Compared to the $N=50$ case, the training process for $N=500$ is noticeably more challenging. In the initial phase, the agent also performs random exploration and experience accumulation, but due to the increased dataset size, the exploration space is larger, requiring more time to accumulate sufficient experience and learn the strategy. After undergoing a rapid improvement in performance, the model continues to steadily enhance its performance until the 5,000th iteration.

The transfer training method starts with significantly higher initial performance compared to models trained from scratch because it leverages the strategies and experiences already learned from the model trained with $N=50$. It can be seen that the performance quickly improves and outperforms the model trained from scratch with $N=500$. Therefore, in our experiments, we use transfer training from the pre-trained model with 50 links for networks with more than 50 links.

\begin{table}[htbp]
\vspace{-5pt}
\centering
\caption{Performance comparison with different inputs on realistic channel model}
\label{Table_realistic_channel}
\begin{tabular}{|c|ccc|}
\hline
\diagbox{$N$}{Input} & distance & CSI & CSI gen.\\ \hline
50  & 0.814          & 1.022              & 1.022                                                                          \\ \hline
100 & 0.803          & 1.025              & 1.023 
\\ \hline
200 & 0.782          & 1.019              & 1.015                                                                          \\ \hline
500 & 0.770          & 1.001              & 0.995                                                                          \\ \hline
\end{tabular}
\end{table}

\subsection{Discussion on Realistic Channels}
In the previous sections, our proposed model and previous data-driven models, such as SpatialLinQ\cite{spatial} and GELinQ\cite{GNNLinQ}, only considered the path loss component of the channel. These models did not require explicit CSI and instead used distance as a proxy, achieving competitive results. This success can be partly attributed to the fact that the path loss based channels are determined by distance.

However, a more realistic channel model 
should account for not only path loss but also antenna gain, shadowing, and fast fading in $h_{ij}$ of Eq-(\ref{Eq_rate}). We collectively refer to this as the ``realistic channel model'', consistent with those used in model-based studies. The realistic channel model presents significant challenges for data-driven methods that rely solely on distance. For example, when SpatialLinQ was tested on channel with path loss and Rayleigh fast fading, its performance significantly deteriorated\cite{spatial}. This outcome is understandable because, under such conditions, the channel becomes a stochastic function of distance, complicating the neural network's objective function.

To adapt our proposed model to the realistic channel, we use CSI as the input, and use the realistic channel gain to compute data rate. In practice, we use the z-score normalization method for CSI, and add an offset to make it non-negative. The results presented in Table 
\ref{Table_realistic_channel} demonstrate the average sum rate ratio of GRLinQ on different network sizes when using two types of input: distance and CSI. Here, ``CSI gen.'' refers to the model trained only with $N=50$ but tested across various network sizes to evaluate its generalizability. It can be seen that for networks utilizing the realistic channel model, merely using the transmitter-receiver distance as inputs is insufficient, achieving only about 80\% of the sum rate relative to FPLinQ, where the latter always has access to exact CSI. If CSI is provided as input to GRLinQ, our model can achieve superior performance. Additionally, even with a significant increase in network density, the experiments demonstrate that our model exhibits strong generalization when CSI is used as input.

\begin{figure}[tbp]
\begin{center}
{\includegraphics[width=0.48\textwidth]{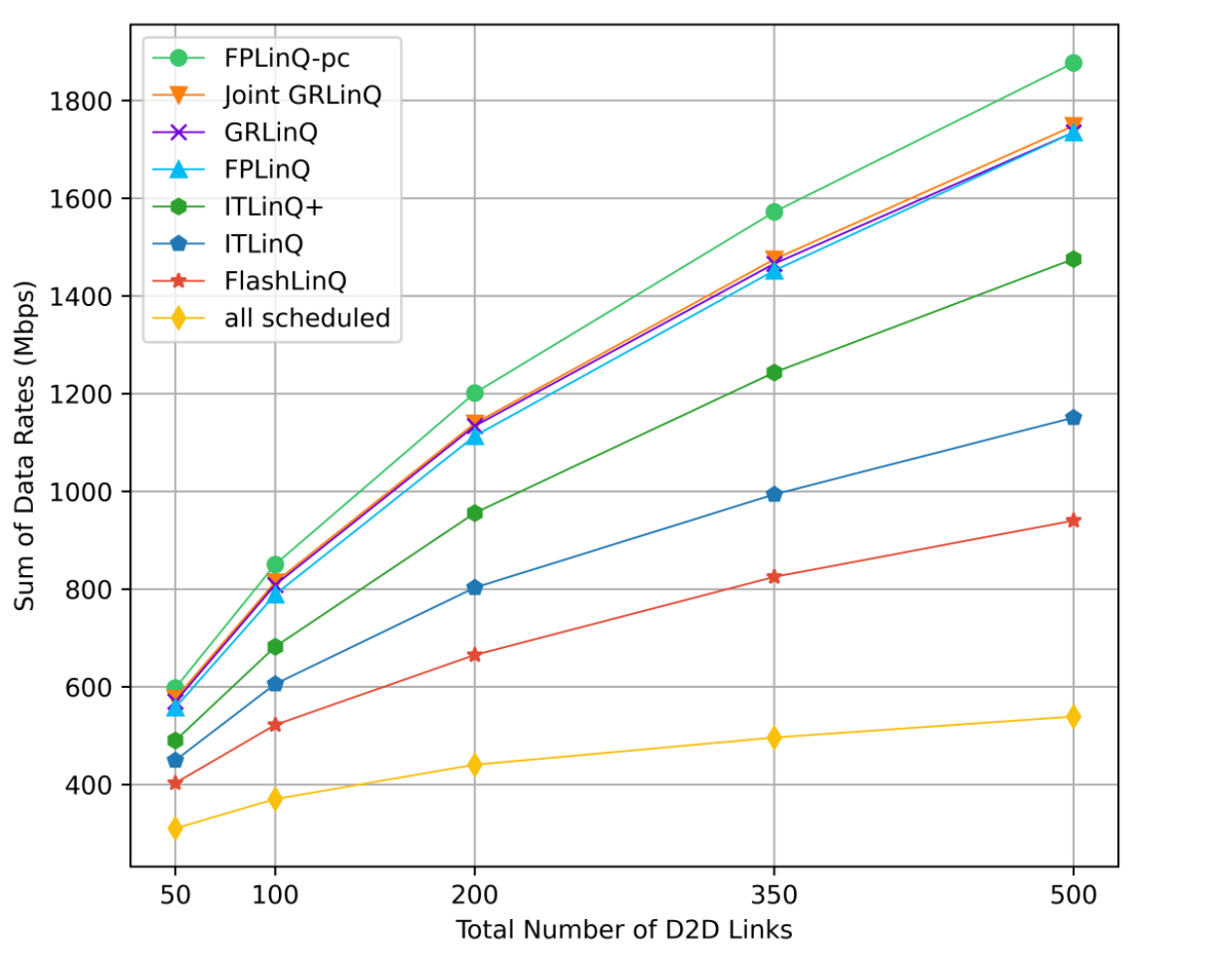}}
\end{center}
   \caption{Sum rate achieved by different approaches on realistic channels.}
   \label{real_channel_joint}
\end{figure}

Finally, we compare the GRLinQ framework with multiple models on a large-scale realistic channel. We additionally introduce a method called Joint GRLinQ, which first uses GRLinQ to generate link scheduling results and then fine-tunes using GRLinQ-pc. As shown in Fig. \ref{real_channel_joint}, both GRLinQ and Joint GRLinQ outperform FPLinQ across all network sizes, and significantly surpass ITLinQ+, ITLinQ, and FlashLinQ. This demonstrates that the GRLinQ framework has a strong capability for solving real channel spectrum sharing. However, there is still a gap between them and FPLinQ-pc, which remains a future direction for improvement.

\section{Conclusion}\label{Conclusion}
This work proposed a novel hybrid model/data-driven intelligent D2D link scheduling  (GRLinQ) and a power control (GRLinQ-pc) mechanism leveraging graph reinforcement learning and experts' knowledge from information theory.
GRLinQ demonstrates superior sum rate performance compared to the optimization-based FPLinQ, with significantly reduced CSI requirements, and to existing data-driven machine learning approaches for link scheduling and/or power control with substantially improved scalability and generalization performance. 
Further studies include theoretical analysis of the convergence/generalization performance of GRLinQ and its potential deployment in practical dynamic wireless networks.





\IEEEtriggeratref{100}

\bibliographystyle{IEEEtran}
\bibliography{paper}


\end{document}